\documentclass[aps, prl, twocolumn, notitlepage, superscriptaddress]{revtex4-1}
\usepackage[T1]{fontenc}
\usepackage{amsmath}
\usepackage{hyperref}
\usepackage{booktabs}
\usepackage{appendix}
\usepackage[inline]{enumitem}
\usepackage{multirow}
\usepackage{amssymb}
\usepackage{amsthm}
\usepackage{array}
\usepackage{graphicx}
\usepackage{verbatim}
\usepackage{bbm}
\usepackage{color}
\usepackage[normalem]{ulem}

\hypersetup{
           breaklinks=true,   
           colorlinks=false,   
           pdfusetitle=true,  
        }
\DeclareGraphicsExtensions{.png,.pdf,.eps}
\renewcommand\Re{\operatorname{Re}}

\def \diracspacing {0.7pt}
\newcommand{\bra}[1]{\langle #1 \hspace{\diracspacing} |} 
\newcommand{\ket}[1]{| \hspace{\diracspacing} #1 \rangle} 
\newcommand{\braket}[2]{\langle #1 \hspace{\diracspacing} | \hspace{\diracspacing} #2 \rangle} 
\newcommand{\ketbra}[2]{| \hspace{\diracspacing} #1 \rangle \langle #2 \hspace{\diracspacing} |} 
\newcommand{\ketbraq}[1]{\ketbra{#1}{#1}} 
\newcommand{\prlparagraph}[1]{\emph{#1.---}}

\newcommand{\abs}[2][]{#1| #2 #1|}

\newcommand{\I}{\mathbb{I}}

\newcommand{\cL}{\mathcal{L}}

\newcommand{\cH}{\mathcal{H}}

\newcommand{\bC}{\mathbb{C}}
\newcommand{\bR}{\mathbb{R}}

\newcommand{\tpsi}{\tilde{\psi}}
\newcommand{\tA}{\tilde{A}}
\newcommand{\tB}{\tilde{B}}

\newcommand{\hx}{\hat{x}}
\newcommand{\hy}{\hat{y}}
\newcommand{\hP}{\hat{P}}
\newcommand{\hQ}{\hat{Q}}

\newcommand{\aux}{\text{aux}}
\newcommand{\ee}{\mathrm{e}}
\newcommand{\ii}{\mathrm{i}}
\DeclareMathOperator{\tr}{tr}

\theoremstyle{definition}
\theoremstyle{plain}

\theoremstyle{remark}

\begin{document}
\title{Unbounded device-independent quantum key rates from arbitrarily small non-locality}
\author{M\'at\'e Farkas}
\email{mate.farkas@york.ac.uk}
\affiliation{Department of Mathematics, University of York, Heslington, York, YO10 5DD, United Kingdom}

\begin{abstract}
Device-independent quantum key distribution allows for proving the security of a shared cryptographic key between two distant parties with potentially untrusted devices. The security proof is based on the measurement outcome statistics (correlation) of a Bell experiment, and security is guaranteed by the laws of quantum theory. While it is known that the observed correlation must be Bell non-local in order to prove security, recent results show that Bell non-locality is in general not sufficient for standard device-independent quantum key distribution. In this work, we show that conversely, there is no lower bound on the amount of non-locality that is sufficient for device-independent quantum key distribution. Even more so, we show that from certain correlations that exhibit arbitrarily small non-locality, one can still extract unbounded device-independent key rates. Therefore, a quantitative relation between device-independent key rates and Bell non-locality cannot be drawn in general. Our main technique comprises a rigorous connection between self-testing and device-independent quantum key distribution, applied to a recently discovered family of Bell inequalities with arbitrarily many measurement outcomes.
\end{abstract}

\maketitle

\prlparagraph{Introduction}
Device-independent quantum key distribution (DIQKD) allows two distant parties to establish a secure cryptographic key without having to trust the devices they use in the protocol \cite{ABG+07,PAB+09,DIQKD_review1,DIQKD_review2}. The security of the key is guaranteed solely by the laws of quantum physics. DIQKD solves two problems present in other types of key distribution protocols: it does not rely either on computational assumptions (like most non-quantum key distribution schemes \cite{RSA78,Mil86,Kob87}), or on the characterisation of the devices used in the protocol (like standard quantum key distribution schemes \cite{ZFQ+08,LWW+10,GLL+11}).
While the practicality of DIQKD still poses challenges, the first proof-of-principle experiments were carried out recently \cite{experiment1,experiment2,experiment3}, demonstrating that DIQKD can be achieved with current technology. Remaining challenges include increasing the key rates and the distance over which the protocols can be implemented, noting that increasing the key rates naturally leads to an increase in the achievable distance as well \cite{DIQKD_review1,DIQKD_review2}.

This work is concerned with precisely these challenges, in particular, with key rates. The key rate of a protocol is the number of secret bits that can be produced in a given round of the protocol, and we will compare key rates with \textit{Bell non-locality} \cite{BCP+14}, a naturally connected notion: in a DIQKD protocol, two parties measure a bipartite quantum system locally, and the final key is extracted from the measurement outcomes. It is known that DIQKD is possible only if these measurement outcome statistics demonstrate non-local correlations (i.e.~they violate a Bell inequality) \cite{ABG+07,PAB+09}. It is, however, less clear how the amount of non-locality (or Bell inequality violation) relates to the achievable key rate. In fact, recently it was shown that non-locality in itself is not sufficient for the security of a large class of DIQKD protocols \cite{FBJL+21}. That is, there exist correlations that violate a Bell inequality, but cannot be used for DIQKD using standard techniques. In this work, we show a somewhat opposing statement: one can extract unbounded key rates from certain correlations that violate Bell inequalities arbitrarily weakly. Furthermore, these protocols also only use standard techniques. Therefore, one can conclude that (standard) DIQKD key rates and Bell non-locality are incomparable resources.

\prlparagraph{Preliminaries}
Any DIQKD protocol starts with the measurement stage: two parties, Alice and Bob, locally measure their part of a fresh copy of a bipartite quantum state $\rho$ defined on the tensor product of two Hilbert spaces, $\cH_A \otimes \cH_B$. Every time they measure, it constitutes a round of the protocol. In every round, they can decide to perform one of (finitely) many available measurements. For Alice, these measurement settings are denoted $x \in \{0,1,\ldots,n_A-1\} =: [n_A]$, and for Bob, $y \in [n_B]$. In each round they obtain a measurement outcome, labelled by $a \in [k_A]$ for Alice and $b \in [k_B]$ for Bob. Once they have obtained their respective outcomes, a new round begins with a fresh copy of $\rho$. They perform many rounds (in this work we are interested in the asymptotic limit of infinitely many rounds) and record their settings and outcomes, which concludes the measurement stage.

After this, Alice and Bob estimate the \emph{correlation}, that is, the joint probability distribution $p(a,b|x,y)$ of the outcomes conditioned on the measurement settings. The estimation is done by publicly announcing a small subset of their inputs and outputs, after which this subset of their data is discarded.

Quantum theory dictates that the correlation will be given by the Born rule,
\begin{equation}
p(a,b|x,y) = \tr[ (A^x_a \otimes B^y_b ) \rho],
\end{equation}
where $\{A^x_a\}_a$ and $\{B^y_b\}_b$ represent positive-operator-valued measures (POVMs) for every $x$ and $y$. That is, $A^x_a$ are positive semidefinite operators on $\cH_A$ such that $\sum_{a \in [k_A]} A^x_a = \I_A$, the identity operator on $\cH_A$, for all $x \in [n_A]$. Similarly, $B^y_b$ are positive semidefinite operators on $\cH_B$ such that $\sum_{b \in [k_B]} B^y_b = \I_B$ for all $y \in [n_B]$.

It is important to note at this point that certain quantum correlations are \emph{non-local} \cite{BCP+14}. Non-local correlations cannot be written in the form
\begin{equation}\label{eq:local}
p_\cL(a,b|x,y) = \sum_\lambda p_{\Lambda}(\lambda) p_A(a|x,\lambda) p_B(b|y,\lambda),
\end{equation}
where $\lambda$ is usually referred to as a \emph{hidden variable}, and $p_\Lambda(\, \cdot \,)$, $p_A(\, \cdot \,|x,\lambda)$ and $p_B(\, \cdot \,|y,\lambda)$ are probability distributions. Correlations of the form \eqref{eq:local} are called \emph{local correlations} (and are considered ``classical'' in the context of Bell non-locality), and they form a convex polytope $\cL$ (for fixed $n_A, n_B, k_A$, and $k_B$). Since $\cL$ is convex, non-local correlations can be witnessed by separating hyperplanes, usually called \emph{Bell inequalities} in this context. A (non-trivial) Bell inequality is a linear functional on the correlations $p(a,b|x,y)$ such that its maximal value attained by local correlations can be exceeded by certain quantum correlations. In such a case, we say that the quantum correlation \emph{violates a Bell inequality}, and is therefore non-local. It is an important pre-requisite for a correlation to violate a Bell inequality in order for it to be useful for DIQKD \cite{ABG+07,PAB+09}.

Returning to the steps of a DIQKD protocol, once they have estimated the correlation, Alice and Bob decide whether the correlation is satisfactory for DIQKD (based on criteria that we will discuss next). If it is not, they abort the protocol. If the correlation passes the test, they employ \emph{privacy amplification} and \emph{error correction} on their remaining data (on the recorded inputs and outputs that were not discarded) \cite{CK78,AC93,Maurer93}. This is done via public, but authenticated classical communication channels. At the end of the privacy amplification and error correction stage, Alice and Bob are each left with a string of bits that are perfectly random (as a result of privacy amplification) to any potential eavesdropper limited by the laws of quantum physics. Moreover, these strings of bits are exactly the same for Alice and Bob, as a result of error correction. The asymptotic key rate, $r$, is then defined as the length of this bit string, divided by the number of rounds, taking the limit of infinitely many rounds.

One of the seminal results of (device-independent) quantum key distribution is a universal lower bound on the achievable key rate from a given correlation. The bound quantifies the key rate that can be extracted from the outcomes of the ``key settings'' $\hx$ on Alice's side and $\hy$ on Bob's side, by performing privacy amplification and error correction via one-way communication from Alice to Bob. The bound is referred to as the Devetak--Winter rate \cite{DW05}, and in our context it is given by
\begin{equation}\label{eq:DW}
r \ge H(A|E) - H(A|B),
\end{equation}
where $H(A|E) = \inf_{\ket{\psi}, \{A^x_a\}, \{B^y_b\} } \{ H(A|E)_\sigma \}$ is the infimum over all states $\ket{\psi} \in \cH_A \otimes \cH_B \otimes \cH_E$ that are a purification of a state $\rho$ on $\cH_A \otimes \cH_B$, and over all POVMs $A^x_a$ on $\cH_A$ and $B^y_b$ on $\cH_B$ such that the state and the measurements are compatible with the observed correlation,
\begin{equation}
\tr[ (A^x_a \otimes B^y_b ) \rho] = p(a,b|x,y).
\end{equation}
Furthermore, $H(A|E)_\sigma$ is the conditional von Neumann entropy of the corresponding classical-quantum state
\begin{equation}
\sigma_{AE} = \sum_{a \in [k_A]} \ketbraq{a} \otimes \tr_{AB}[ (A^{\hx}_a \otimes \I_B \otimes \I_E) \ketbraq{\psi}],
\end{equation}
and $H(A|B)$ is the conditional Shannon entropy of the distribution $p(a,b|\hx,\hy)$. Note that an analogous bound holds for the case of one-way communication from Bob to Alice, and that the bound on $r$ only depends on the observed correlation $p(a,b|x,y)$. Furthermore, this bound is valid against the most powerful, so-called coherent eavesdropping attacks \cite{VV14,Arn20,ADF+18}. If Alice and Bob cannot establish a positive lower bound for their key rate, they abort the protocol.

The term $H(A|E)$ captures the cost of privacy amplification: to model an eavesdropper (Eve) limited by quantum physics, we consider a tripartite quantum state on some Hilbert space $\cH_A \otimes \cH_B \otimes \cH_E$ such that Eve holds the subsystem $E$. Then, $H(A|E)$ quantifies Eve's uncertainty of Alice's outcome of measurement $\hx$, given her quantum side-information in the worst-case scenario, optimised over the possible tripartite states and measurements compatible with the observed correlation. In other words, $H(A|E)$ quantifies the \emph{device-independent (private) randomness} of Alice's outcome of her measurement~$\hx$.

The term $H(A|B)$ captures the cost of error-correction, since it quantifies Bob's uncertainty of Alice's outcome of measurement $\hx$ given his classical information, which is the outcome of his measurement $\hy$. Since $H(A|B)$ can be directly computed from the correlation, the difficulty in estimating the Devetak--Winter bound is in bounding $H(A|E)$. Indeed, various methods have been proposed to bound this quantity for arbitrary correlations. General analytic techniques are lacking, while numerical techniques scale rather badly in the number of settings and outcomes \cite{BSS14,NSPS14,TSG+21,BFF21,BFF23}. It is important to note, however, that once we have a bound on $H(A|E)$, the lower bound on the key rate can be boosted by finding an appropriate measurement $\hy$ for Bob that results in small conditional entropy $H(A|B)$.

\prlparagraph{Self-testing and DIQKD}
In this work, we analytically tackle the problem of bounding private randomness from a specific type of correlations. In particular, we expose a rigorous connection between DIQKD and a strong certification technique in Bell non-locality called \emph{self-testing} \cite{SB20}. We say that a correlation $p(a,b|x,y)$ self-tests the quantum state $\ket{\tpsi} \in \cH_{\tA} \otimes \cH_{\tB}$ and the measurements $\tA^x_a$ on $\cH_{\tA}$ and $\tB^y_b$ on $\cH_{\tB}$, if for all quantum states $\rho$ on some Hilbert spaces $\cH_A \otimes \cH_B$ and all measurements $A^x_a$ on $\cH_A$ and $B^y_b$ on $\cH_B$ such that $p(a,b|x,y) = \tr[ (A^x_a \otimes B^y_b) \rho ]$, we have that for every purification $\ket{\psi} \in \cH_A \otimes \cH_B \otimes \cH_E$ of $\rho$ there exist Hilbert spaces $\cH_{A'}$ and $\cH_{B'}$ and local isometries $V_A: \cH_A \to \cH_{\tA} \otimes \cH_{A'}$ and $V_B: \cH_B \to \cH_{\tB} \otimes \cH_{B'}$ such that
\begin{equation}\label{eq:self-test}
(V_A \otimes V_B \otimes \I_E) (A^x_a \otimes B^y_b \otimes \I_E) \ket{\psi} = (\tA^x_a \otimes \tB^y_b) \ket{\tpsi} \otimes \ket{\aux}
\end{equation}
for some state $\ket{\aux} \in \cH_{A'} \otimes \cH_{B'} \otimes \cH_{E}$ and for all $x,y,a,b$. The primary aim of self-testing is to characterise quantum states and measurements solely from the observed correlation in a Bell experiment, and many examples of self-testing are known \cite{SB20}.

Our main technical observation linking self-testing and DIQKD is that whenever a correlation self-tests some quantum state and measurements, the parties can extract private randomness from their measurements. In fact, it is sufficient that the weaker form of Eq.~\eqref{eq:self-test},
\begin{equation}\label{eq:self-test_weaker}
(V_A \otimes V_B \otimes \I_E) (A^{\hx}_a \otimes \I_B \otimes \I_E) \ket{\psi} = (\tA^{\hx}_a \otimes \I_{\tB}) \ket{\tpsi} \otimes \ket{\aux}
\end{equation}
holds for some fixed $\hx$ and for all $a$. Eq.~\eqref{eq:self-test_weaker} follows from Eq.~\eqref{eq:self-test} by fixing $x = \hx$ and summing up over $b$, and note that further summing up over $a$, we obtain
\begin{equation}\label{eq:self-test_weaker2}
(V_A \otimes V_B \otimes \I_E)  \ket{\psi} =  \ket{\tpsi} \otimes \ket{\aux}.
\end{equation}

The reason why condition \eqref{eq:self-test} is sufficient for certifying private randomness is because the term $H(A|E)$ in Eq.~\eqref{eq:DW} can be computed analytically if the correlation is self-testing (or the weaker condition \eqref{eq:self-test_weaker} holds) as follows. For all tripartite states $\ket{\psi}$ and measurements $A^x_a$ compatible with the observed correlation, we have
\begin{widetext}
\begin{equation}
\begin{split}
\sigma_{AE} & \left. = \sum_a \ketbraq{a} \otimes \tr_{AB}\left[ (A^{\hx}_a \otimes \I_B \otimes \I_E) \ketbraq{\psi} \right] = \sum_a \ketbraq{a} \otimes \tr_{AB}\left[ (V^\dagger_A V_A \otimes V^\dagger_B V_B \otimes I_E) (A^{\hx}_a \otimes \I_B \otimes \I_E) \ketbraq{\psi} \right] \right. \\
& \left. = \sum_a \ketbraq{a} \otimes \tr_{\tA A' \tB B'}\left[ (V_A \otimes V_B \otimes I_E) (A^{\hx}_a \otimes \I_B \otimes \I_E) \ketbraq{\psi} (V^\dagger_A \otimes V^\dagger_B \otimes I_E)\right] \right. \\
& \left. = \sum_a \ketbraq{a} \otimes \tr_{\tA A' \tB B'}\left[ \left( (\tA^{\hx}_a \otimes \I_{\tB}) \ket{\tpsi}_{\tA \tB} \otimes \ket{\aux}_{A'B'E} \right) \left( \bra{\tpsi}_{\tA \tB} \otimes \bra{\aux}_{A'B'E} \right) \right] \right. \\
& \left. = \sum_a \ketbraq{a} \otimes \tr \left[ (\tA^{\hx}_a \otimes \I_{\tB}) \ketbraq{\tpsi}_{\tA \tB} \right] \tr_{A'B'} \left( \ketbraq{\aux}_{A'B'E} \right) = \left[ \sum_a p_A(a|\hx) \ketbraq{a} \right] \otimes \sigma_E,
\right.
\end{split}
\end{equation}
\end{widetext}
where $\sigma_E = \tr_{A'B'} \ketbraq{\aux}_{A'B'E}$ is some fixed quantum state on $\cH_E$, $p_A(a|\hx) = \sum_b p(a,b|\hx,y)$ is Alice's marginal distribution, and we used the conditions \eqref{eq:self-test_weaker} and~\eqref{eq:self-test_weaker2}. The conditional von Neumann entropy in Eq.~\eqref{eq:DW} is then given by (for all states and measurements compatible with the correlation)
\begin{equation}
\begin{split}
H(A|E)_\sigma & \left. = H(AE)_\sigma - H(E)_\sigma \right. \\
& \left. = H \left( \sum_a p_A(a|\hx) \ketbraq{a} \otimes \sigma_E \right) - H(\sigma_E) \right. \\
& \left. = H \left( \sum_a p_A(a|\hx) \ketbraq{a} \right) + H\left( \sigma_E \right) - H(\sigma_E) \right. \\
& \left. = H\left( \{ p_A(a|\hx) \}_a \right) = H(A),
\right.
\end{split}
\end{equation}
where we used that the von Neumann entropy is additive under the tensor product. Therefore, the entropy $H(A)$ of Alice's outcome from measurement $\hx$ is private, that is, no eavesdropper can guess it better than random. Note that a similar argument is used in the proofs of Ref.~\cite{WBC22}.

In order to promote this device-independent randomness certification statement to device-independent quantum key distribution, we need a measurement on Bob's side such that its outcome is correlated with the outcome of setting $\hx$ of Alice. Such a choice maximises the Devetak--Winter bound in Eq.~\eqref{eq:DW} by minimising $H(A|B)$. While such a highly correlated measurement setting $\hy$ might already be part of the setup that gives rise to the self-testing correlation, notice that adding an extra setting on Bob's side does not change the calculation for Alice's private randomness. Therefore, for every self-testing correlation one can aim to find the best possible measurement for Bob that maximises the device-independent key rate. Note once again that this maximisation is much less cumbersome than computing $H(A|E)$, since $H(A|B)$ can be directly computed from the observed correlation $p(a,b|x,y)$. Therefore, given $\ket{\tilde{\psi}}$ and $\{ \tA^{\hx}_a \}_a$ from the self-testing statement, one can attempt to find a measurement $\{ B^{\hy}_b \}_b$ minimising $H(A|B)$.

\prlparagraph{Unbounded key from arbitrarily small non-locality}
Using the above techniques, we will now prove that from correlations arbitrarily close to the local set (and therefore violating any Bell inequality arbitrarily weakly) one can extract $\log(d)$ bits of device-independent key for any integer $d \ge 2$. 
For this purpose, we use the Bell inequalities from Ref.~\cite{PAK22}. The inequalities are parametrised by an integer $d \ge 2$ and and \emph{overlap matrix} $O$, whose elements are characterised by two orthonormal bases on $\bC^d$, which we choose to be $\{ \ket{j} \}_{j=0}^{d-1}$ and $\{ \ket{e_k} \}_{k=0}^{d-1}$. The elements of the overlap matrix are then given by
\begin{equation}
O_{jk} = \abs{ \braket{j}{e_k} }.
\end{equation}
In the Bell scenario, Alice has 2 measurement settings with $d$ outcomes each and Bob has $d^2$ settings with 3 outcomes each (notice that we swapped the role of Alice and Bob compared to Ref.~\cite{PAK22}).
For every $d \ge 2$ and every overlap matrix such that $O_{jk} < 1$ for all $j,k$ (equivalently, $O_{jk} > 0$ for all $j,k$) the authors of Ref.~\cite{PAK22} construct a non-trivial Bell inequality, i.e.~a Bell inequality that has a quantum violation. Moreover, they show that the maximal quantum violation can be achieved by sharing a locally $d$-dimensional maximally entangled state, $\ket{\phi^+_d} = \frac{1}{\sqrt{d}} \sum_{j=1}^d \ket{jj}$ and Alice's measurements being $\{ \ketbraq{j} \}_{j=0}^{d-1}$ for $x=1$ and $\{ \ketbraq{e_k} \}_{k=0}^{d-1}$ for $x=2$.

While the maximal violation of these inequalities does not provide a self-test in the usual sense, in Ref.~\cite{PAK22} it is shown that for every state $\rho$ on $\cH_A \otimes \cH_B$ giving rise to the maximal violation, there exist local isometries $V_A: \cH_A \to \bC^d \otimes \cH_{A'}$ and $V_B: \cH_B \to \bC^d \otimes \cH_{B'}$ (with $\cH_{A'} \simeq \cH_A$ and $\cH_{B'} \simeq \cH_B$) such that
\begin{equation}\label{eq:isometry_maxent}
( V_A \otimes V_B ) \rho (V_A^\dagger \otimes V_B^\dagger) = \ketbraq{ \phi^+_d } \otimes \sigma_{A'B'}
\end{equation}
for some quantum state $\sigma_{A'B'}$ on $\cH_{A'} \otimes \cH_{B'}$.
Furthermore, it can be shown from the maximal violation \footnote{See the Supplemental Material, which includes Refs.~\cite{GKW+18,Laura}} that Alice's measurement corresponding to setting $x=1$ satisfies
\begin{equation}\label{eq:isometry_meas}
V_A A^1_a V_A^\dagger = \ketbraq{a} \otimes \tilde{A} \quad \forall a \in [d]
\end{equation}
for some fixed operator $\tilde{A}$ on $\cH_{A'}$ and
\begin{equation}\label{eq:isometry_B}
V_B V_B^\dagger = \I_{\bC^d} \otimes \tilde{B},
\end{equation}
for some fixed operator $\tilde{B}$ on $\cH_{B'}$. By an argument similar to the one in the previous section, it can be shown~\cite{Note1} using Eqs.~\eqref{eq:isometry_maxent}, \eqref{eq:isometry_meas}, and \eqref{eq:isometry_B} that for every purification $\ket{\psi}_{ABE}$ of $\rho$, the resulting conditional von Neumann entropy will again satisfy
\begin{equation}
H(A|E)_{\sigma} = H(A) = \log(d)
\end{equation}
for the setting $\hx = 1$ of Alice [the specific value $\log(d)$ follows from the fact that $p_A(a|1) = \tr[ (A^1_a \otimes \I_B )\rho ] = \tr[ ( \ketbraq{a} \otimes \I_B ) \ketbraq{ \phi^+_d } ] = \frac1d$ due to the relations \eqref{eq:isometry_maxent} and \eqref{eq:isometry_meas}].
Then, introducing a measurement for Bob that is perfectly correlated to the $\hx=1$ setting of Alice (e.g.~in the ideal realisation one can choose $B^{\hy}_b = \ketbraq{b}$), we get a lower bound on the key rate,
\begin{equation}
r \ge H(A|E) - H(A|B) = \log(d)
\end{equation}
for all $d \ge 2$ and for all overlap matrices with $O_{jk} > 0$. That is, we obtain a family of correlations that certify $\log(d)$ bits of secret key. Notice that while these correlations maximally violate a Bell inequality, in some cases they might be arbitrarily close to the set of local correlations.

Exploiting precisely this fact,
we now show that for every $d \ge 2$, there exist correlations arbitrarily close to the local set, but still certifying $\log(d)$ bits of secret key. To do so, we need to provide an overlap matrix $O$ such that the correlation maximising the corresponding Bell inequality from Ref.~\cite{PAK22} is arbitrarily close to the local set. Consider the trivial case of $\ket{e_k} = \ket{k}$ for all $k \in [d]$, leading to an overlap matrix $O_{jk} = \delta_{jk}$. The corresponding correlation that arises by measuring $\ket{\phi^+_d}$ with the measurements $\{ \ketbraq{j} \}$ for both settings $x=1$ and $x=2$ is local, since Alice's measurements are compatible (they are equal) \cite{QVB14}. Now let us perturb $\{ \ket{e_k} \}$ by a small unitary transformation in a way that leads to a non-trivial overlap matrix with all elements strictly positive. One particularly symmetric way to achieve this is by taking the generalised Pauli X operator
\begin{equation}
X = \sum_{j=0}^{d-1} \ketbra{j+1}{j} = \sum_{j=0}^{d-1} \omega_d^j \ketbraq{ \chi_j } = \ee^G,
\end{equation}
where $\omega_d = \ee^{\frac{2 \pi \ii }{d}}$ is the $d$-th root of unity, $\ket{\chi_j} = \frac{1}{\sqrt{d}}\sum_{k=0}^{d-1} \omega_d^{jk}\ket{k}$ is the Fourier basis and $G = \sum_{j=0}^{d-1} \left( \frac{2 \pi \ii }{d} j \right) \ketbraq{ \chi_j }$. Then, consider the unitary operator parametrised by $\varepsilon \in [0,1]$,
\begin{equation}
U_\varepsilon := \sum_{j=0}^{d-1} \omega_d^{\varepsilon j} \ketbraq{ \chi_j } = \ee^{\varepsilon G}.
\end{equation}
Clearly, $U_0 = \I$, and $U_\varepsilon$ is continuous in $\varepsilon$, a consequence of the well-known fact that the map $t \mapsto \ee^{t M}$ is continuous in $t$ for any matrix $M$ (see e.g.~\cite[Chapter 2]{hall2013lie}). Also, for every $\varepsilon \in (0,1)$ we have that the overlap matrix of $\{ \ket{j} \}_{j=0}^{d-1}$ and $\{ U_\varepsilon \ket{k} \}_{k=0}^{d-1}$ is non-trivial, that is, all of its elements are strictly positive \cite{Note1}. As such, by the above arguments, the correlation that arises by measuring $\ket{\phi^+_d}$ with the measurements $\{ \ketbraq{j} \}_{j=0}^{d-1}$ and $\{ U_\varepsilon \ketbraq{k} U_\varepsilon^\dagger \}_{k=0}^{d-1}$ (and the appropriate measurements for Bob) certifies $\log(d)$ bits of secure key for every $\varepsilon \in (0,1)$ and every integer $d \ge 2$.

If we now choose $\varepsilon$ to be arbitrarily small (but positive), the resulting correlation, $p_\varepsilon(a,b|x,y)$ gets arbitrarily close (e.g.~in $\ell_1$ norm) to the local correlation $p_{\varepsilon = 0}(a,b|x,y)$, since the correlation is also continuous in~$\varepsilon$ (it is quadratic in $U_\varepsilon = \ee^{\varepsilon G}$). Therefore, for any integer $d \ge 2$, for arbitrarily small $\varepsilon > 0$ the correlation $p_\varepsilon(a,b|x,y)$ certifies $\log(d)$ bits of device-independent key, but the correlation is arbitrarily close to the set of local correlations. That is, from arbitrarily small non-locality, one can still certify unbounded (with increasing~$d$) device-independent key.

\prlparagraph{Conclusion}
In this work, we exposed a rigorous connection between self-testing and DIQKD as well as device-independent randomness generation. Thanks to this connection and the latest developments in high-dimensional Bell non-locality, we showed that unbounded device-independent key rates can be certified from correlations with arbitrarily small non-locality. This result together with recent findings indicates that DIQKD and Bell non-locality might be incomparable resources, and in the search for a fundamental quantum resource for DIQKD, the amount of non-locality is not the right quantity to consider.

\prlparagraph{Note added}
During the completion of this work, the author became aware of the related independent work of Ref.~\cite{WBC23}. The authors there derive new self-testing statements in the simplest Bell scenario and prove that constant DIQKD rates (1 bit) can be achieved from arbitrarily small non-locality.

\prlparagraph{Acknowledgements}
The author would like to thank J\k{e}drzej Kaniewski and Laura Man\v{c}inska for fruitful discussions.

\vspace*{0.4cm}
\bibliography{bib_simple_attack}

\newpage

\onecolumngrid

\appendix

\vspace{1cm}

\begin{center}
\large{\bf Supplemental Material}
\end{center}

\section{The local isometries}

In this section, we prove Eqs.~\eqref{eq:isometry_meas} and \eqref{eq:isometry_B} from the main text. In order to do so, we introduce some notation, following Ref.~\cite{PAK22}. We denote the operators corresponding to Alice's setting $x=1$ by $\{ P_a \}_{a=0}^{d-1}$ and those corresponding to setting $x=2$ by $\{ Q_a \}_{a=0}^{d-1}$. In Ref.~\cite{PAK22}, the authors show that the maximal violation of the Bell inequalities corresponding to a given overlap matrix $O$ certifies that Alice's measurements satisfy
\begin{align}\label{eq:PQP}
P_j Q_k P_j = O_{jk}^2 P_j \\ \label{eq:QPQ}
Q_k P_j Q_k = O_{jk}^2 Q_k 
\end{align}
for all $j,k \in [d]$. This also implies that both $\{ P_a \}$ and $\{Q_a\}$ are projective measurements, by summing over the middle index in the above relations. The isometry on Alice's side is then given by $V_A : \cH_A \to \bC^d \otimes \cH_{A'}$
\begin{equation}
V_A = \frac{1}{ O_{d-1,j} } \sum_{k=0}^{d-1} \frac{1}{ O_{kj} } \ket{k} \otimes P_{d-1} Q_j P_k
\end{equation}
for some fixed (but arbitrary) $j$, and such that $\cH_{A'} \simeq \cH_A$. Then, we have
\begin{equation}
\begin{split}
V_A P_a V_A^\dagger & \left. = \left( \frac{1}{ O_{d-1,j} } \sum_{k=0}^{d-1} \frac{1}{ O_{kj} } \ket{k} \otimes P_{d-1} Q_j P_k \right) P_a \left( \frac{1}{ O_{d-1,j} } \sum_{\ell=0}^{d-1} \frac{1}{ O_{\ell j} } \bra{\ell} \otimes P_{\ell} Q_j P_{d-1} \right) \right. \\
& \left. = \frac{1}{ O_{d-1,j}^2 }\sum_{k,\ell =0}^{d-1} \frac{1}{ O_{kj} O_{\ell j}} \ketbra{k}{\ell} \otimes P_{d-1} Q_j P_k P_a P_\ell Q_j P_{d-1} = \frac{1}{ O_{d-1,j}^2 O_{aj}^2} \ketbraq{a} \otimes P_{d-1} Q_j P_a Q_j P_{d-1} \right. \\
& \left. = \frac{1}{ O_{d-1,j}^2 } \ketbraq{a} \otimes P_{d-1} Q_j P_{d-1} = \ketbraq{a} \otimes P_{d-1},
\right.
\end{split}
\end{equation}
where we used that $P_k P_a P_\ell = \delta_{k,a} \delta_{\ell,a} P_a$ and the relations in Eqs.~\eqref{eq:PQP} and \eqref{eq:QPQ}. This proves Eq.~\eqref{eq:isometry_meas} in the main text.

In order to prove Eq.~\eqref{eq:isometry_B} in the main text, we recall from Ref.~\cite{PAK22} that the isometry on Bob's side is given by $V_B : \cH_B \to \bC^d \otimes \cH_{B'}$
\begin{equation}
V_B = \frac{1}{ O_{d-1,j} } \sum_{k=0}^{d-1} \frac{1}{ O_{kj} } \ket{k} \otimes \hP_{d-1} \hQ_j \hP_k
\end{equation}
where $\hP_j$ and $\hQ_k$ are operators on $\cH_{B'} \simeq \cH_B \simeq \cH_A$ satisfying the same relations as the operators of Alice, that is, the relations in Eqs.~\eqref{eq:PQP} and \eqref{eq:QPQ}, and $j$ is again fixed but arbitrary. Then, we have
\begin{equation}
\begin{split}
V_B V_B^\dagger & \left. = \left( \frac{1}{ O_{d-1,j} } \sum_{k=0}^{d-1} \frac{1}{ O_{kj} } \ket{k} \otimes \hP_{d-1} \hQ_j \hP_k \right) \left( \frac{1}{ O_{d-1,j} } \sum_{\ell=0}^{d-1} \frac{1}{ O_{\ell j} } \bra{\ell} \otimes \hP_{\ell} \hQ_j \hP_{d-1} \right) \right. \\
& \left. = \frac{1}{ O_{d-1,j}^2 }\sum_{k,\ell =0}^{d-1} \frac{1}{ O_{kj} O_{\ell j}} \ketbra{k}{\ell} \otimes \hP_{d-1} \hQ_j \hP_k \hP_\ell \hQ_j \hP_{d-1} = \frac{1}{ O_{d-1,j}^2} \sum_{k=0}^{d-1} \frac{1}{ O_{kj}^2} \ketbraq{k} \otimes \hP_{d-1} \hQ_j \hP_k \hQ_j \hP_{d-1} \right. \\
& \left. = \frac{1}{ O_{d-1,j}^2 } \sum_{k=0}^{d-1} \ketbraq{k} \otimes \hP_{d-1} \hQ_j \hP_{d-1} = \sum_{k=0}^{d-1} \ketbraq{k} \otimes \hP_{d-1} = \I_{\bC^d} \otimes \hP_{d-1},
\right.
\end{split}
\end{equation}
where we used that $\hP_k \hP_\ell = \delta_{kl} \hP_k$ and the relations \eqref{eq:PQP} and \eqref{eq:QPQ} for $\hP_j$ and $\hQ_k$. This proves Eq.~\eqref{eq:isometry_B} in the main text.

\section{Privacy from maximal Bell inequality violation}

In this section, we show that the conditions in Eqs.~\eqref{eq:isometry_maxent}, \eqref{eq:isometry_meas} and \eqref{eq:isometry_B} are sufficient for certifying the secrecy of Alice's outcome. First, we have that for every state $\rho$ on $\cH_A \otimes \cH_B$ and every measurement $A^1_a$ that give rise to the maximal violation of the corresponding Bell inequality, we have
\begin{equation}\label{eq:self-test_matrix}
(V_A \otimes V_B) (A^1_a \otimes \I_B) \rho (V_A^\dagger \otimes V_B^\dagger) = ( \ketbraq{a} \otimes \I_{\bC^d} ) \ketbraq{ \phi^+_d } \otimes (\tilde{A} \otimes \tilde{B}) \sigma_{A'B'}
\end{equation}
as a direct consequence of Eqs.~\eqref{eq:isometry_maxent}, \eqref{eq:isometry_meas} and \eqref{eq:isometry_B}. Since the trace of the left-hand side is the marginal probability of Alice, $p(a|1) = \frac1d$, and the operator on the left-hand side is positive semidefinite, we have that $(\tilde{A} \otimes \tilde{B})\sigma_{A'B'}$ is a quantum state.

Eq.~\eqref{eq:self-test_matrix} is reminiscent of the ``matrix form'' self-testing \cite{GKW+18} with the difference that Bob's measurement operators do not appear. It can be shown that this matrix form condition is equivalent to the vector form condition \cite{Laura}, that is, Eq.~\eqref{eq:self-test_matrix} is equivalent to
\begin{equation}
(V_A \otimes V_B \otimes \I_E)( A^1_a \otimes \I_B \otimes \I_E) \ket{\psi}_{ABE} = (\ketbraq{a} \otimes \I_{\bC^d}) \ket{ \phi^+_d} \otimes \ket{\aux}_{A'B'E}
\end{equation}
for all purifications $\ket{\psi}_{ABE}$ of $\rho$, for some $\ket{\aux}_{A'B'E}$. This condition is exactly of the form of Eq.~\eqref{eq:self-test_weaker} in the main text, and therefore the secrecy of Alice's outcome from the measurement setting $x=1$ follows by the arguments in the main text.

\section{Non-trivial overlap matrix from the perturbed basis}

In this section we show that the overlap matrix
\begin{equation}
O_{jk} = \abs{ \bra{j} U_\varepsilon \ket{k} }
\end{equation}
satisfies $O_{jk} > 0$ for all $j,k$ whenever $\varepsilon \in (0,1)$. We have
\begin{equation}
\bra{j} U_\varepsilon \ket{k} = \bra{j} \sum_{\ell=0}^{d-1} \omega_d^{\varepsilon \ell} \ket{\chi_\ell} \braket{\chi_\ell}{k} = \sum_{\ell=0}^{d-1} \omega_d^{\varepsilon \ell} \frac{1}{ \sqrt{d} } \omega_d^{-\ell k} \braket{j}{\chi_\ell} = \frac1d \sum_{\ell = 0}^{d-1} \omega_d^{( \varepsilon - k + j ) \ell} = 
\begin{cases}
1 & \text{if } \varepsilon - k + j = 0 \\
\frac1d \frac{1 - \omega_d^{( \varepsilon - k + j ) d } }{1 - \omega_d^{ \varepsilon - k + j  } } & \text{otherwise,}
\end{cases}
\end{equation}
where in the last step we used the formula for a geometric sum. Notice that for $\varepsilon \in (0,1)$, the condition $\varepsilon - k + j = 0$ cannot hold, so we can always consider the second case. As such, using the fact that for any $\alpha \in \bR$ we have
\begin{equation}
\abs{ 1 - \omega_d^\alpha }^2 = (1 - \omega_d^\alpha)(1 - \omega_d^{-\alpha}) = 2 - \omega_d^\alpha - \omega_d^{-\alpha} = 2(1 - \Re( \omega_d^\alpha ) ) = 2 \left[ 1 - \cos \left( \frac{2\pi}{d} \alpha \right) \right],
\end{equation}
we obtain
\begin{equation}
0 \le \abs{ \bra{j} U_\varepsilon \ket{k} }^2 = \frac{1}{d^2} \cdot \frac{ 1 - \cos\left[ 2 \pi (\varepsilon - k + j) \right]} { 1 - \cos\left[ \frac{2 \pi}{d} (\varepsilon - k + j) \right]}
\end{equation}
and equality holds if and only if $\cos\left[ 2 \pi (\varepsilon - k + j) \right] = 1$, that is, if and only if $(\varepsilon - k + j) \in \mathbb{N}$, which cannot hold if $\varepsilon \in (0,1)$. Therefore, we conclude that $\abs{ \bra{j} U_\varepsilon \ket{k} }^2 > 0$, as required.

\end{document}